\documentclass{article}
\usepackage{amsmath,amsfonts,amssymb}

\newcommand\mfk\mathfrak
\newcommand\mcl\mathcal
\newcommand\mbb\mathbb
\newcommand\mtl\mathit
\newcommand\mbf\mathbf
\newcommand\onm\operatorname
\newcommand\op[1]{\mathord{\mbox{\bf\itshape #1}}}
\newcommand{\set}[2]{\bigl\{\mskip1mu #1:#2\mskip1mu\bigr\}}
\newcommand{\sset}[1]{\bigl\{\mskip1mu#1\mskip1mu\bigr\}}
\newcommand{\llangle}{\langle\!\langle}\newcommand{\rrangle}{\rangle\!\rangle}

\begin{document}
\title{Lagrangian mechanics without ordinary differential equations}
\author{George W. Patrick}
\date{September 2005}
\maketitle
{\renewcommand{\thefootnote}{}
\footnotetext{$^*$Partially supported by the
Natural Sciences and Engineering Research Council, Canada.}}
\begin{abstract}
A variational proof is provided of the existence and uniqueness
of evolutions of regular Lagrangian systems.
\end{abstract}

\section*{Introduction}
Let $\mcl Q$ be a smooth, finite dimensional manifold and $L\colon
T\mcl Q\rightarrow\mbb R$ be a smooth Lagrangian. Evolutions of
the Lagrangian system defined by $L$ are by definition 
the $C^1$ curves $q\colon[0,h]\rightarrow\mbb R$ which are critical
points of the action
\begin{equation*}
S_h=\int_0^hL\circ q^\prime(t)\,dt,
\end{equation*}
subject to the constraint that $q(0)$ and $q(h)$ are constant. A
typical route to existence and uniqueness (given that $L$ is regular)
of the Lagrangian evolutions, is to to show that derivatives
$q^\prime(t)$ of evolution curves $q(t)$ are integral curves of the 
Lagrangian vector field $X_E$, constructed either using the
Euler-Lagrange equations in charts, or using the Lagrange two-form
$\omega_L$, the energy $E$, and the equation
\begin{equation*}
\op i_{X_E}\omega_L=\op dE.
\end{equation*}
In any case, standard ODE theory provides existence and uniqueness of
the initial value problem $q(0)=q_0,q^\prime(0)=v_0$.  For a
self-contained exposition, see~\cite{AbrahamRMarsdenJE-1978-1}.

Given two nearby $q_1,q_2\in \mcl Q$, does there exist a unique
evolution curve $q(t)$ such that $q(0)=q_1$ and $q(h)=q_2$?
This is the local boundary value problem of Lagrangian mechanics.  The
problem crops up in a variety of situations. For example:
\begin{enumerate}
\item 
If $\mcl Q$ is a Riemannian manifold and $L(v)=\frac12 g(v,v)$, then
the Lagrangian evolution curves are constant speed reparameterizations
of the geodesics, and the local boundary value problem becomes that of
locating the unique local geodesic connecting two sufficiently nearby
points.
\item  
A solution to the local boundary value problem is required to
construct type~1 generating functions $S_t(q_2,q_1)$ for the
Hamiltonian flow, which are defined by
\begin{equation*}
S_t(q_2,q_1)=\int_0^tL\circ q(t)\,dt
\end{equation*}
where $q(t)$ is the evolution curve with $q(0)=q_1$, $q(t)=q_2$.
\end{enumerate}

After constructing the Lagrangian flow $F^{X_E}_t$, the solution to
the local boundary value problem is obtained by solving the equations
\begin{equation*}
\tau_{\mcl Q}F^{X_E}_t(v_{q_1})=q_2
\end{equation*}
for $v_{q_1}\in T_{q_1}\mcl Q$ as a function of $q_1,q_2,t$, where
$\tau_{\mcl Q}\colon T\mcl Q\rightarrow \mcl Q$ is the canonical
projection. This may appear to be a straightforward application of the
implicit function theorem near $t=0,q_1=q_2$, but that is not quite
so, because the equation fails to be appropriately regular there. With
some care, however, the local boundary value problem can be solved by
this route~\cite{PatrickGW-1991-1}.

But, first solving the initial value problem seems like a rather
circuitous route to the solution of the local boundary value problem,
especially considering that the boundary values $q_1$ and $q_2$
actually occur as the constraints in the original variational
formulation for the evolution. Regular constrained optimization
problems have critical points which persist as smooth functions of the
constraint values.  Why not solve the boundary value problem directly,
avoiding an excursion into the initial value problem via ODE theory?

The obstruction to simply getting on with the job is a fundamental
one: \emph{the problem of finding the critical points of $S_h$ subject
to the constraint $q(0)=q_0$, $q(t)=q_1$, is nonregular at $h=0$,
which precisely where one wants to perturb from.} Indeed, the
objective function $S_h$ is actually zero when $h=0$. More seriously,
the constraint
\begin{equation*}
q(t)\mapsto \bigl(q(0),q(h)\bigr)
\end{equation*}
maps into the diagonal of $\mcl Q\times\mcl Q$ at $h=0$, and hence
cannot be, even formally, a submersion. One cannot perturb from such a
degenerate landscape.

In this article, I provide a direct variational proof of the local
boundary value problem, using a regularization procedure, which is
adapted from the one used in~\cite{CuellC-PatrickGW-2005-2} for a
similar problem in the context of discrete Lagrangian systems. The
regularization procedure culminates in the replacement of the
variational problem with an equivalent, regular one, after which
readily available techniques used to prove the infinite dimensional
Morse Lemma, and the (infinite dimensional) implicit function theorem,
give the result.

\section{Regularization}
Assume that $L\colon\mcl Q\rightarrow\mbb R$ is $C^r$ with $r\ge2$.
Since the aim is to provide a local result, also assume that $\mcl Q$
is an open subset of $\mbb R^n$, and that $\bar q\in \mcl Q$ is given.
One seeks a perturbative approach from $t=0$, $q_1=q_2=\bar q$, so
that $q_1$ and $q_2$ will be near $\bar q$.  The regularization
procedure is, step-by-step, as follows:

\bigskip\noindent1.~\emph{Transform the variational problem from one
for curves in $\mcl Q$ to one for curves in $T\mcl Q$, with an additional first
order constraint.} The transformation is simply to seek critical
points $\bigl(q(t),v(t)\bigr)\in T\mcl Q$ of the objective
\begin{equation*}
S_h=\int_0^hL\bigl(q(t),v(t)\bigr)\,dt,
\end{equation*}
subject to the constraints
\begin{equation*}
v(t)=\frac{dq}{dt},\quad q(0)=q_1,\quad q(h)=q_2.
\end{equation*}
There are additional freedoms inherent in the use of
curves in $T\mcl Q$ rather than curves in $\mcl Q$, and these will be important
for the regularization.

\medskip\noindent2.~\emph{Reparameterize, so that the solutions
curves, which are defined on $[0,h]$, do not disappear as
$h\rightarrow0^+$}. The curves
$\bigl(q(t),v(t)\bigr)$, $t\in[0,h]$ are replaced by the curves
$\bigl(Q(t),V(t)\bigr)$, $t\in[0,1]$ through
\begin{equation*}
Q(u)=q(hu),\quad V(u)=v(hu),\qquad u\in[0,1].
\end{equation*}
For $h>0$, the new curve $\bigl(Q(u),V(u)\bigr)$ satisfies an
equivalent variational problem, which can be worked out as
follows. First, substitute $u=t/h$ and divide by $h$ to obtain
\begin{equation*}
\frac1h\int_0^hL\bigl(q(t),v(t)\bigr)\,dt=\int_0^1L\bigl(Q(u),V(u)\bigr)\,du.
\end{equation*}
Since $h$ is constant for the variational principle, one can use the
right-hand-side as an objective for $\bigl(Q(u),V(u)\bigr)$. The first
order constraint transforms as follows:
\begin{equation*}
\frac d{dt}q(t)-v(t)
=\left(\frac1h\frac d{du}\bigl(Q(u)\bigr)-V(u)
  \right)_{u=\frac th}.
\end{equation*}
Multiplying this by $h$, the reparameterized variational principle is
for curves $\bigl(Q(u),V(u)\bigr)$ with values in $T\mcl Q$ which are
critical points of the objective
\begin{equation*}
S=\int_0^1L\bigl(Q(u),V(u)\bigr)\,du,
\end{equation*}
subject to the constraints
\begin{equation*}
\frac d{du}\bigl(Q(u)\bigr)-hV(u)=0,\quad Q(0)=q_1,\quad Q(1)=q_2.
\end{equation*}
Notice that both the objective and the constraints are smooth through
$h=0$ and that, at $h=0$, the first order constraint is equivalent to
the constraint that the curve $\bigl(Q(u),V(u)\bigr)$ lies in a
fiber of $T\mcl Q$ i.e. the curve is vertical.

\medskip\noindent3.~\emph{Restrict the remaining (boundary)
constraints to the submanifold defined by the regularized first order
constraint, and regularize the result.} The first order constraint may
be solved smoothly through $h=0$ by integration:
\begin{equation}\label{10}
Q(u)=Q(0)+h\int_0^u V(s)\,ds.
\end{equation}
Thus the set of curves $\bigl(Q(u),V(u)\bigr)$ may be replaced by
the set of curves $\sset{V(u)}$. Also, $q_2$ may be replaced by
$q_1+hz$, $z\in\mbb R^n$, and then, since under~\eqref{10},
\begin{equation*}
Q(1)=Q(0)+h\int_0^1 V(s)\,ds,
\end{equation*}
the objective and the boundary constraint become, respectively,
\begin{equation}\label{20}
S_h=\int_0^1L\left(q_1+h\int_0^u V(s)\,ds,V(u)\right)\,du,\quad
\int_0^1V(u)\,du=z,
\end{equation}
while $h$ and $q_1$ appear as parameters. 

This completes the procedure, since the variational problem~\eqref{20}
is formally regular through $h=0$. Indeed, at $h=0$,~\eqref{20}
becomes finding the curves $V(u)$ which are critical points of the
constrained problem
\begin{equation*}
S_0=\int_0^1L\bigl(q_1,V(u)\bigr)\,du,\quad
\int_0^1V(u)\,du=z.
\end{equation*}
Using a Lagrange multiplier $\lambda$, the solutions are given by
setting
\begin{equation*}
\int_0^1\frac{\partial L}{\partial v}\bigl(q_1,V(u)\bigr)\delta V(u)\,du=
\int_0^1\lambda\cdot\delta V(u)\,du
\end{equation*}
for all $\delta V(u)$ i.e.
\begin{equation*}
\frac{\partial L}{\partial v}\bigl(q_1,V(u)\bigr)=\lambda.
\end{equation*}
If $L$ is a regular Lagrangian then this implies $V(u)$ is constant, and the
constraint then implies $V(u)=z$. Thus the solution to~\eqref{20} at
constraint value $z$ is the constant curve $z$, and there is exactly
one critical point for each constraint value.  

\section{Implicit function theorem solution to the 
regularized variational problem}

Consider $S_h$ from~\eqref{20} on the Banach space $C^k([0,1],\mbb
R^n)$ of curves $V(u)$, where $0\le k\le r$. Since $L$ is $C^r$, the
Omega Lemma~(\cite{AbrahamR-MarsdenJE-RatiuTS-1988-1},~page~102)
implies that the integrand of $S_h$ is $C^r$ as a map into
$C^0([0,1],\mbb R)$. Since integration on $C^0([0,1],\mbb R)$ is
bounded linear (and therefore $C^\infty$), it follows that $S_h$ is
$C^r$, irrespective of the value of~$k$.

I specialize the method for proving the infinite dimensional Morse
lemma~\cite{GolubitskyM-MarsdenJE-1983-1,TrombaAJ-1976-1,TrombaAJ-1983-1}
to the constrained variational problem~\eqref{10}.  One first
calculates the gradient of $S$ from the derivative $\op dS_h$ using
the $C^2([0,1],\mbb R^n)$ weak inner product
\begin{equation*}
\llangle V,W\rrangle=\int_0^1 V\cdot W.
\end{equation*}
That computation is as follows (for short, below $Q(u)$ means the
right side of~\eqref{10}):
\begin{align*}
&\op dS_h([V(u)])\delta V(u)\\
&\qquad
  =\frac d{d\epsilon}\int_0^1L\left(q_1+h\int_0^u V(s)+\epsilon\delta V(s)
  \,ds,V(u)+\epsilon\delta V(u)\right)\,du\\
&\qquad
  =\int_0^1\left(\frac{\partial L}{\partial q}
  \bigl(Q(u),V(u)\bigr)h\int_0^u\delta V(s)\,ds
+\frac{\partial L}{\partial v}\bigl(Q(u),V(u)\bigr)\delta V(u)\right)\,du\\
&\qquad
  =\int_0^1\int_s^1h\frac{\partial L}{\partial q}\bigl(Q(u),V(u)\bigr)
  \delta V(s)\,du\,ds
  +\int_0^1\frac{\partial L}{\partial v}\bigl(Q(u),V(u)\bigr)\delta V(u)\,du\\
&\qquad
  =\int_0^1\int_u^1h\frac{\partial L}{\partial q}\bigl(Q(s),V(s)\bigr)
  \delta V(u)\,ds\,du
  +\int_0^1\frac{\partial L}{\partial v}\bigl(Q(u),V(u)\bigr)\delta V(u)\,du\\
&\qquad
  =\int_0^1\left(
  \int_u^1h\frac{\partial L}{\partial q}\bigl(Q(s),V(s)\bigr)\,ds
  +\frac{\partial L}{\partial v}\left(Q(u),V(u)\right)\right)\delta V(u)\,du,
\end{align*}
from which
\begin{equation*}
\nabla S_h=
  \frac{\partial L}{\partial v}\bigl(Q(u),V(u)\bigr)
  +h\int_u^1\frac{\partial L}{\partial q}\bigl(Q(s),V(s)\bigl)\,ds,
\quad
Q(u)=q_1+h\int_0^u V(s)\,ds.
\end{equation*}
By the same reasoning as was used to find the differentiability of
$S_h$, the gradient $\nabla S_h$ (best thought of as a vector field) is
a $C^{r-1}$ map from $C^k([0,1],\mbb R^n)$ to $C^0([0,1],\mbb R^n)$,
also irrespective of the value of $k$, $0\le k\le r-1$.

The constraint of~\eqref{20} (i.e. the second equation) is $C^\infty$
because it is bounded linear, and its derivative is
\begin{equation*}
\delta V(u)\mapsto\int_0^1\delta V(u)\,du.
\end{equation*}
The kernel of this derivative, say $\mbb E_0$, is the tangent space
to the constraint set, and it splits $C^k([0,1],\mbb R^n)$
orthogonally with respect to the metric $\llangle,\rrangle$ (the
complement is the subspace of constant functions) by
\begin{equation*}
\delta V(u)=\left(\delta V(u)-\int_0^1\delta V(u)\right)
\oplus\int_0^1\delta V(u)
\end{equation*}
$S_h$ has a critical point on the level sets of the constraint if and
only if the orthogonal projection $\mbb P_{\mbb E_0}$ of $\nabla S_h$
to the kernel $\mbb E_0$ is zero i.e. for the solutions to the
constrained variational problem~\eqref{20}, one solves
\begin{equation*}
\mbb P_{\mbb E_0}\nabla S_h(V_0\oplus V_1)=0
\end{equation*}
for $V_0$ near
\begin{equation*}
V_0=0,\quad V_1=0,\quad z=0,\quad q_1=\bar q,\quad h=0.
\end{equation*}
To use the implicit function theorem, one requires that the
appropriate partial derivative of $\mbb P_{\mbb E_0}\nabla S_h$ is a
linear isomorphism. Remembering to set $h=0$, that derivative is
\begin{align*}
\left.\frac d{d\epsilon}\right|_{\epsilon=0}
\mbb P_{\mbb E_0}\frac{\partial L}{\partial v}\bigl(\bar q,
  \epsilon\delta V_0(u)\bigr)
  &=\mbb P_{\mbb E_0}\frac{\partial^2L}{\partial v^2}(\bar q,0)\delta V_0(u)\\
  &=\frac{\partial^2L}{\partial v^2}(\bar q,0)\delta V_0(u)
  -\int_0^1\frac{\partial^2L}{\partial v^2}(\bar q,0)\delta V_0(u)\\
  &=\frac{\partial^2L}{\partial v^2}(\bar q,0)\delta V_0(u).
\end{align*}
If $L$ is regular, this is a linear isomorphism of $\mbb E_0$, with inverse
\begin{equation*}
 \delta V_0(u)\mapsto\left(\frac{\partial^2L}{\partial v^2}(\bar q,0)
 \right)^{-1}\delta V_0(u).
\end{equation*} 
Thus, the implicit function theorem provides neighborhoods
$W_1\subseteq\mbb R^n\times\mbb R^n\times\mbb R=\sset{(q_1,z,h)}$
containing $(\bar q,0,0)$ and $W_2\subseteq C^k([0,1],\mbb R^n)$ of
the constant map $u\mapsto 0$, and a $C^{r-1}$ map $\psi\colon
W_1\rightarrow W_2$ such that for all $(q_1,z,h)\in W_1$,
$\psi(q_1,z,h)\in C^k([0,1],\mbb R^n)$ is the unique critical point in
$W_2$ of the constrained variational problem~\eqref{20}. By setting
$k=0$ and then $k=r-1$, one can arrange that $W_2$ is a $C^{r-1}$
neighborhood, $\psi$ has values in $W_2$, and hence in the $C^{r-1}$
curves, $\psi$ is $C^{r-1}$ with the $C^{r-1}$ topology, but that
$\psi$ provides the unique solution among the $C^0$ curves in a $C^0$
open neighborhood, say
\begin{equation*}
\set{V(u)}{|V(u)|<\epsilon},
\end{equation*}
of the constant curve $0$.

Now reverse the regularization.  Pick an $h>0$ such that $(\bar
q,0,h)\in W_1$, set
\begin{equation*}
\bar W_1=\set{(q_1,q_2)}{\bigl(q_1,(q_2-q_1)/h,h)\in W_1}
\end{equation*}
and define
\begin{equation*}
\bar\psi_{(q_1,q_2)}(t)=
q_1+\int_0^{\frac{t}{h}}\psi\left(q_1,\frac1h(q_2-q_1),h\right)(u)\,du.
\end{equation*}
Then $(\bar q,\bar q)\in\bar W_1$, $\bar\psi_{(q_1,q_2)}(t)$ is
defined for $(q_1,q_2)\in\bar W_1$ and $t\in[0,h]$, and
$\bar\psi_{(q_1,q_2)}(t)$ is a first order curve in $T\mcl Q$ which
has base integral curve a Lagrangian evolution. This evolution is
unique among the continuous curves corresponding to $|V(u)|<\epsilon$
i.e. among $C^1$ curves $q(t)$ such that $|q^\prime(t)|<\epsilon/h$,
so $C^1$ curves $q(t)$ in some $C^1$ neighborhood of the constant curve
$\bar q$.

\section{Remarks}

The regularization can be formulated in invariant terms on the
manifold $\mcl Q$, using a tubular neighborhood of the antisymmetric
normal bundle of the diagonal of $\mcl Q\times\mcl Q$ to accomplish
the subtraction $q_2-q_1$.  Replacing $\bigl(Q(u),V(u)\bigr)$ with its
$T\mcl Q$ version $V(u)$, the regularized variational problem,
at $h=0$, becomes
\begin{equation*}
S_0=\int_0^1L\circ V,\quad \int_0^1V(u)=z,\quad \tau_{\mcl Q}V=\mbox{constant}.
\end{equation*}
It is a pretty result that the variational principle on $\mcl Q$
regularizes to this trivial one on the fibers of $T\mcl Q$.
The map $\bar\psi$ is defined only for small $z=(q_2-q_1)/h$, and
since $\bar\psi(q_1,q_2)$ is a solution which goes from $q_1$ to $q_2$
in time $h$, the velocity of this solution is also, approximately,
$(q_2-q_1)/h$. Thus regularizing only at $z=0$ provides evolutions
which correspond only to velocities near zero.  This is unacceptable
for the objective of solving the initial value problem by first
solving the local boundary value problem, because it assigns
evolutions only to those initial data corresponding to velocities near
zero, while it is known of course from ODE theory that there is a
unique integral curve of the Lagrangian vector field corresponding to
any velocity.  However, minor extensions of the above show that the
variational principle actually regularizes at all $z$.  The local
solutions so obtained along the entire solution (i.e. $z$ maps to the
constant curve $V(u)=z$) of the regularized variational principle at
$h=0$, may be glued together using a technique the can be found for
example in~\cite{LangS-1972-1},~page~{\sf 97}. This provides solutions
starting at any velocity.  For the same reason, this also an important
step in the discrete Lagrangian
context~\cite{CuellC-PatrickGW-2005-2}, where the discrete initial
value problem is addressed by first solving the local boundary value
problem.

\end{document}